\pdfoutput=1
\documentclass[aps,prl,reprint,10pt,superscriptaddress,longbibliography,amsmath,amssymb]{revtex4-2}
\usepackage{graphicx}
\usepackage{mathtools}
\usepackage{newtxtext}
\usepackage{newtxmath}
\usepackage[colorlinks=true,linkcolor=blue,citecolor=blue,urlcolor=blue]{hyperref}
\usepackage{orcidlink}

\begin{document}
\title{A Universal Law for Horizon Formation in Black Hole Mergers}
\author{Yu-Cun Xie\,\orcidlink{0000-0002-7603-7411}}
\email{yzx5296@psu.edu}
\affiliation{Institute for Gravitation and the Cosmos, The Pennsylvania State University, University Park, PA 16802, USA}
\affiliation{Department of Physics, The Pennsylvania State University, University Park, PA 16802, USA}
\author{Vaishak Prasad\,\orcidlink{0000-0001-6712-2457}}
\affiliation{Institute for Gravitation and the Cosmos, The Pennsylvania State University, University Park, PA 16802, USA}
\affiliation{Department of Physics, The Pennsylvania State University, University Park, PA 16802, USA}
\date{November 11, 2026}

\begin{abstract}
When two black holes merge, a common horizon appears around them as a single surface, which immediately splits into an outer and an inner one. We show that this event has a universal local form. Without any symmetry assumption, the two surfaces separate as $(t-t_*)^{1/2}$ along a single, everywhere positive mode and share a drift linear in $t-t_*$. Three numerical simulations, including an eccentric, precessing one, confirm the exponent to half a percent, and the reconstructed formation surface is marginally stable, as predicted.
\end{abstract}
\maketitle

Of the three stages of a binary black hole coalescence, the inspiral is described by post-Newtonian theory and the ringdown by perturbations of a single black hole. The merger between them, where the gravitational-wave signal peaks \cite{LIGOScientific:2016aoc}, is fully nonlinear and is known only from numerical solutions of Einstein's equations. This Letter shows that one event in that stage, the formation of the horizon of the final black hole, follows a simple law, which we derive.

\begin{figure}[t]
 \begin{minipage}[t]{0.43\columnwidth}
  \raggedright (a)\\[1pt]
  \centering\includegraphics[height=5.8cm,trim=108 18 110 4,clip]{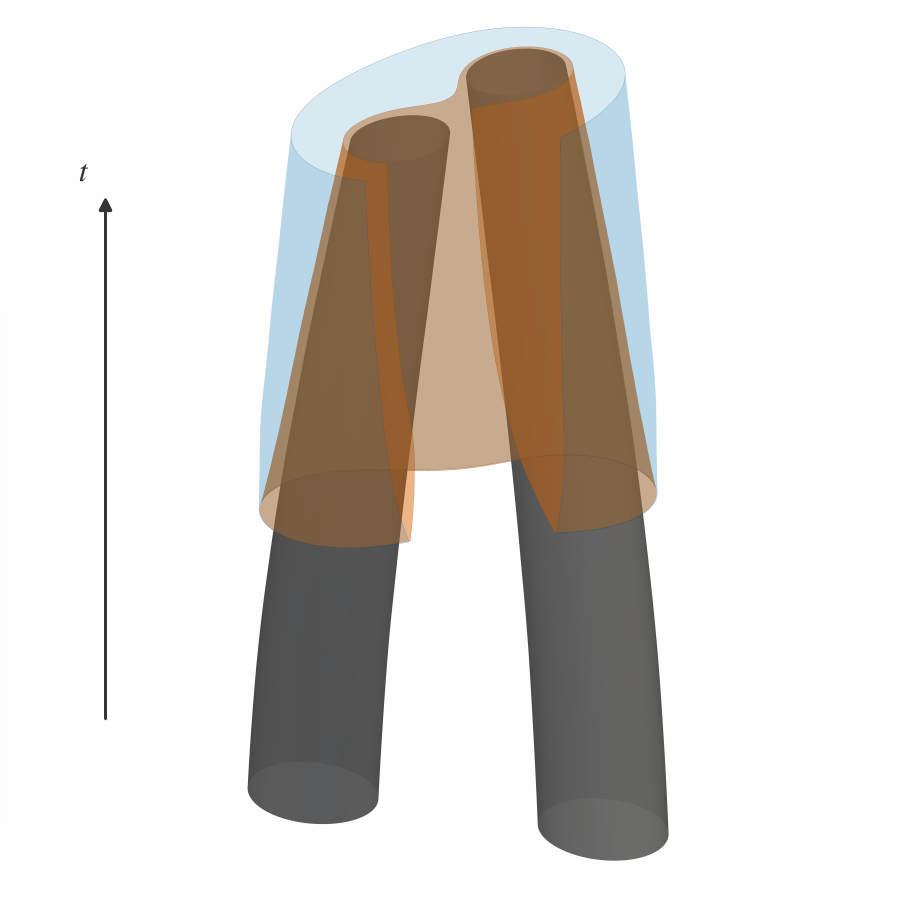}
 \end{minipage}\hfill
 \begin{minipage}[t]{0.55\columnwidth}
  \raggedright (b)\\[1pt]
  \centering\includegraphics[height=5.8cm]{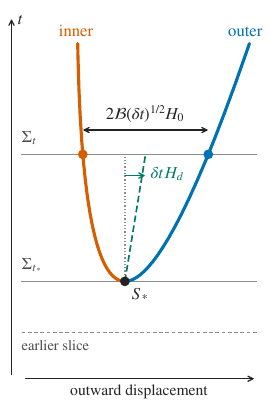}
 \end{minipage}
 \caption{Formation of the common horizon. (a) Horizons of Simulation~I in spacetime, with time running upward and one spatial dimension suppressed; adapted from the companion article \cite{XiePrasad:Companion2026}. The individual horizons (gray) persist inside the common horizon, which is cut open to show them. It appears at $t_*$ and has an outer (blue) and an inner (vermilion) branch. (b) The same structure at one point of $S_*$: one smooth tube that is tangent to the slice $\Sigma_{t_*}$ and is intersected twice by a later slice $\Sigma_t$. The intersections are separated by $2\mathcal B(\delta t)^{1/2}H_0$, and their midpoint has moved by $\delta t\,H_d$.}
 \label{fig:tube}
\end{figure}

On a slice of constant time, a marginally outer trapped surface (MOTS) is a closed surface whose outgoing light rays have zero expansion, and the apparent horizon is the outermost such surface \cite{AshtekarKrishnan:2025recent}. Unlike the event horizon, it can be located without knowing the future. Before the merger each hole has its own apparent horizon. The common horizon does not form by these two joining and deforming into one, the way two soap bubbles merge. Consider instead a surface made of the two horizons joined by a short neck. On the neck the surface is strongly curved inward, so outgoing light rays converge there. Once the holes are close enough, the outgoing light rays of the joined surface are nonexpanding everywhere, and the outermost MOTS must then lie outside it \cite{Andersson:2008up}. Hence, before the two horizons touch, at a time $t_*$, a new MOTS appears around both holes, and on later slices it is found as two surfaces, an outer and an inner common horizon \cite{Pook-Kolb:2018igu,Pook-Kolb:2019iao,Pook-Kolb:2021gsh}, as in Fig.~\ref{fig:tube}(a). The outer one is the horizon of the remnant. The original horizons persist inside, and the inner common horizon moves inward toward them \cite{Pook-Kolb:2019iao}.

\begin{figure}[t]
 \centering
 \includegraphics[width=0.92\columnwidth]{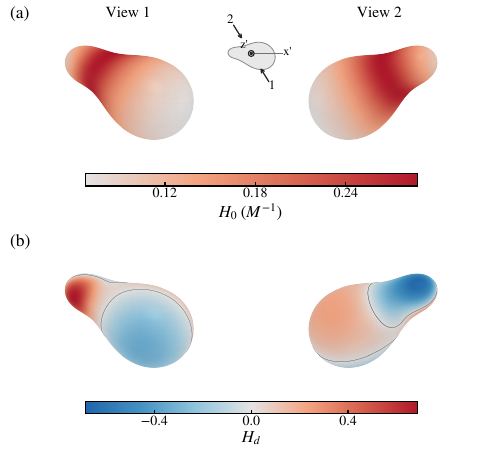}
 \caption{The two fields of Eq.~\eqref{eq:formation-law} on the formation surface $S_*$ of Simulation~III, the eccentric, precessing binary. (a) The positive mode $H_0$, along which the two common horizons separate. (b) The drift $H_d$ shared by both; black curves mark $H_d=0$. Each row shows two opposite views, and the key gives the viewing directions. Both fields are obtained from the first five pairs of horizons \cite{XiePrasad:Companion2026}.}
 \label{fig:fields}
\end{figure}

We find that the two common horizons are displaced from the surface $S_*$ on which they form according to
\begin{equation}
 \delta H_\pm
 =\pm\mathcal B(\delta t)^{1/2}H_0
  +\delta t\,H_d+O((\delta t)^{3/2}) .
 \label{eq:formation-law}
\end{equation}
Here $\delta t=t-t_*$, and $\delta H_\pm$ is the displacement of the outer ($+$) and inner ($-$) horizon from $S_*$ along its outward normal, at each point of $S_*$. The constant $\mathcal B$ is positive, and $H_0$ and $H_d$ are two functions on $S_*$. The square-root term has opposite signs on the two horizons and separates them. Because $H_0$ is positive everywhere, they separate over the whole surface at once, and one lies entirely inside the other. The linear term is shared, and it moves and deforms both horizons together. Figure~\ref{fig:tube}(b) shows both terms at one point of $S_*$. The exponent $1/2$ is the same for every merger, while $\mathcal B$, $H_0$, and $H_d$ depend on the binary and are fixed by the geometry at $S_*$. Figure~\ref{fig:fields} shows the two fields over the whole surface in an eccentric, precessing merger. A companion article gives the complete derivation, numerical setup, and validation \cite{XiePrasad:Companion2026}.

\begin{figure*}[t]
 \centering
 \includegraphics[width=\textwidth]{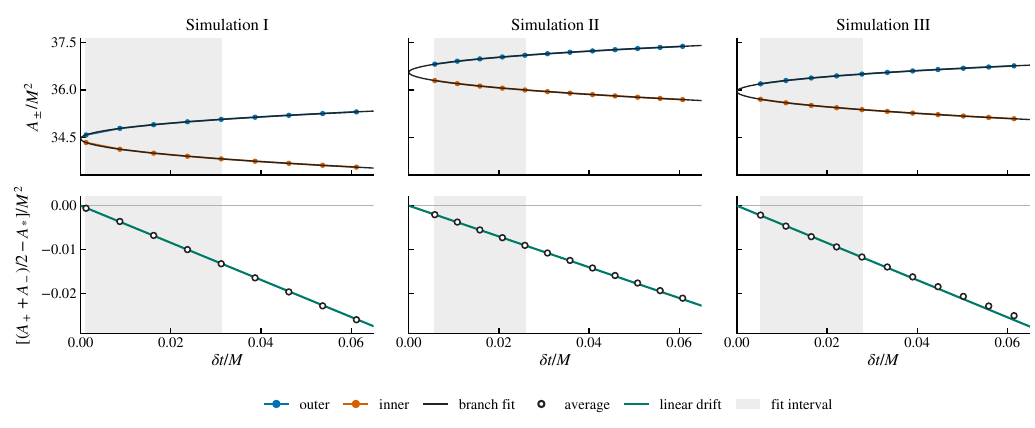}
 \caption{Area of the common horizons in Simulations~I--III. Upper panels: areas of the outer and inner horizons, with fits to Eq.~\eqref{eq:quantity}. Lower panels: mean of the two areas minus the fitted formation area $A_*$; the straight line is the linear term of Eq.~\eqref{eq:quantity}. Shading marks the five pairs used in the fits, and the curves are continued beyond them. In each simulation $\delta t$ is measured from the formation time estimated from the area.}
 \label{fig:area}
\end{figure*}

\textit{Zero mode at formation.}
The law follows because $S_*$ is both the first common MOTS and the outermost one. The expansion $\Theta$ of the outgoing light rays depends on the shape of the surface and on the slice in which it lies. Suppose a small change of shape could always compensate a small change of slice. Then $S_*$ could be continued smoothly to slightly earlier slices as well as to later ones, and it would not be the first. At $S_*$ there must therefore be a change of shape that leaves $\Theta$ unchanged to first order. This is a zero mode $H_0$ of the stability operator, the linear response of $\Theta$ to a change of shape \cite{Andersson:2007fh}.

The stability operator has a real principal eigenvalue $\lambda_0$, the one of smallest real part, and its eigenfunction is positive everywhere \cite{Andersson:2007fh}. A MOTS is stable when $\lambda_0\geq0$, which means that it can be deformed slightly outward with its outgoing expansion remaining zero or positive. The outermost MOTS on a slice is always stable. The zero mode of $S_*$ must then be the principal one. Hence $\lambda_0=0$ at formation, $H_0$ is positive, and there is no second zero mode. The argument rests on the continuation and tangency theorems for MOTSs \cite{Andersson:2005gq,Andersson:2008up}.

\textit{Local branch structure.}
A displacement $\mathcal A H_0$ along the zero mode does not change $\Theta$ at first order, so its amplitude $\mathcal A$ is fixed at second order. At leading order, $\Theta=0$ reduces to a single equation,
\begin{equation}
 c_t\,\delta t+c_2\mathcal A^2=0 .
 \label{eq:reduced}
\end{equation}
The coefficient $c_t$ measures how fast the expansion changes as the slice advances, and $c_2$ measures its second-order response to a displacement along $H_0$. The law requires both to be nonzero, and their signs are then determined: outermost stability gives $c_2>0$, and the absence of a common horizon before $t_*$ gives $c_t<0$ \cite{XiePrasad:Companion2026}. If $c_2$ vanished, the cubic term would dominate and give a single branch with $\mathcal A\propto(\delta t)^{1/3}$. Two branches with exponent $1/2$ therefore indicate a quadratic tangency.

Equation~\eqref{eq:reduced} is a parabola, $\delta t\propto\mathcal A^2$. In spacetime it means that the two common horizons form one smooth tube of MOTSs, and that the time slices are tangent to this tube at $S_*$, where the slice time has a minimum along the tube (Fig.~\ref{fig:tube}). No slice before $t_*$ intersects the tube. Slices just after $t_*$ intersect it twice, at $\mathcal A=\pm\mathcal B(\delta t)^{1/2}$ with $\mathcal B^2=-c_t/c_2$, and the two intersections are the outer and inner common horizons. The geometry is regular at $t_*$.

All other deformations are determined by $\mathcal A$, because the linear response is invertible on the complement of the zero mode. One order further, the same reduction gives the shared drift, with three parts, $H_d=\mathcal C H_0+H_t+\mathcal B^2H_2$: a shift of the amplitude, the response of the surface to the advance of the slice, and its second-order response to the opening \cite{XiePrasad:Companion2026}. In bifurcation theory this structure is called a fold \cite{Kielhofer:2012bt}, and related bifurcations of marginal surfaces are studied in concurrent work by Booth, Cox, and Okpala \cite{BoothCoxOkpala:2026bifurcation}. 

\textit{Dependence on the foliation.}
A null horizon is the same in every foliation, because each of its cross sections is marginally trapped \cite{AshtekarKrishnan:2025recent}, but a non-null tube of MOTSs is not. Where such a tube is spacelike, its foliation by marginally trapped surfaces is unique \cite{AshtekarGalloway:2005uniqueness}, so a different slicing of the same tube does not yield MOTSs. A different foliation of spacetime in general generates a different tube. The structure derived here nevertheless does not single out a foliation. In any foliation in which a tube first appears with $c_t$ and $c_2$ nonzero, its two branches open as $(\delta t)^{1/2}$ with a shared linear drift. Within one foliation, a reparametrization of time rescales $\mathcal B$ and $H_d$ but not the exponent.

\textit{Area and stability.}
Any smooth quantity $\mathcal Q$ measured on the horizons has the same form,
\begin{equation}
 \mathcal Q_\pm-\mathcal Q_*
 =\pm\mathcal A_{\mathcal Q}(\delta t)^{1/2}
  +\mathcal B_{\mathcal Q}\,\delta t+O((\delta t)^{3/2}),
 \label{eq:quantity}
\end{equation}
where $\mathcal Q_*$ is its value on $S_*$. Both coefficients are given by the geometry at $S_*$; for instance, $\mathcal A_{\mathcal Q}$ is $\mathcal B$ times the first variation of $\mathcal Q$ along $H_0$ \cite{XiePrasad:Companion2026}. For the area, the outer horizon gains area and the inner horizon loses the same amount at leading order, and $dA/dt$ diverges at $t_*$. The area is nevertheless smooth along the tube; the divergence arises because the slice time is stationary along the tube at $S_*$. The values of $\mathcal Q$ on the two horizons are one smooth function of the tube parameter, evaluated at $\pm(\delta t)^{1/2}$. Their difference therefore contains only odd powers of $(\delta t)^{1/2}$ and their mean only integer powers of $\delta t$, as Fig.~\ref{fig:area} shows for the area. The squared separation is then a regular function of time, which gives $t_*$ before $S_*$ is known.

The principal eigenvalue of each horizon has the same form, starting from $\lambda_0=0$, with square-root coefficient $2c_2\mathcal B/\langle H_0^\dagger|H_0\rangle$, where $H_0^\dagger$ is the zero mode of the adjoint operator and the bracket is the integral of their product over $S_*$ \cite{XiePrasad:Companion2026}. It is positive on the outer horizon, which is stable, and negative on the inner one, which is unstable from the moment it forms.

\textit{Numerical simulations.}
We test these predictions in the three binary black hole evolutions of the companion article \cite{XiePrasad:Companion2026}, performed with the Einstein Toolkit \cite{Loffler:2011ay}, using puncture initial data \cite{Ansorg:2004ds} and the moving-puncture method \cite{Campanelli:2005dd,Baker:2005vv}, whose lapse condition defines the time slices. Simulations~I and II are nonspinning, quasicircular binaries with mass ratios $1$ and $2$. Simulation~III has mass ratio $2$, an eccentric orbit, and spins of $0.4$ in the orbital plane, so it precesses and has no symmetry. We use units of $M$, the sum of the initial black hole masses. Common horizons are located with AHFinderDirect \cite{Thornburg:2003sf}, and their area and shear are computed with QuasiLocalMeasures \cite{SchnetterKrishnanBeyer:2006dynamicalNR}.

The zero mode makes the Newton iteration of the horizon finder singular at $t_*$, so the formation surface itself cannot be found directly. We reconstruct $t_*$ from the squared separation of the first five pairs of horizons found after formation, which span $\delta t\lesssim0.03M$, and $S_*$ from their mean \cite{XiePrasad:Companion2026}.

On the reconstructed $S_*$, with the geometry interpolated to $t_*$, the expansion vanishes to numerical accuracy and the principal eigenvalue is about $10^5$ times smaller than the next one \cite{XiePrasad:Companion2026}, as predicted for the formation surface.

Measured from $S_*$, half the difference of the two horizons is, to leading order, $\mathcal B(\delta t)^{1/2}H_0$, and their mean is $\delta t\,H_d$. As predicted, the measured $H_0$ is positive everywhere in all three simulations. It is largest in the neck between the holes, while $H_d$ changes sign (Fig.~\ref{fig:fields}).

\begin{figure}[t]
 \centering
 \includegraphics[width=\columnwidth]{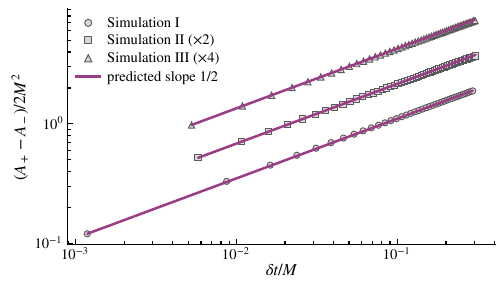}
 \caption{The square-root law in the area. Half the difference between the areas of the outer and inner horizons against the time from formation, on logarithmic axes. Simulations~II and III are shifted upward for clarity. The purple lines are the predicted slope $1/2$, drawn through the first point of each simulation with no other adjustable parameter.}
 \label{fig:sqrt}
\end{figure}

In Fig.~\ref{fig:sqrt}, half the difference of the two areas follows a line of slope $1/2$ on logarithmic axes out to $\delta t\simeq0.3M$, ten times the extent of the fitted window. When the exponent is left free in fits to the first five pairs, with the next-order term included, it is within half a percent of $1/2$ for the surface geometry, the area, and the integrated squared shear of the outgoing light rays. The companion gives all fits, their dependence on the fitting interval, and the resolution studies \cite{XiePrasad:Companion2026}.

\textit{Discussion.}
A horizon finder converges poorly as $\lambda_0\to0$, so detection of the first common horizon can be delayed. In Simulation~II the fitted $t_*$ precedes the first detected pair by more than one interval between searches, so the finder missed a slice on which both horizons existed. The fit recovers the formation time and surface from later, well-conditioned data.

This analysis applies at a generic later tangency between a tube of MOTSs and the time slices, where interior horizons appear or annihilate in pairs \cite{Pook-Kolb:2021gsh}, with the same square-root law. A pair that annihilates is the time reverse of the fold. There the zero mode need not be the principal one, and the two surfaces need not be nested. Joining such local descriptions along the tubes may give an analytic account of the horizon structure inside a merger.

The drift $H_d$ contains the coordinate motion and deformation common to both horizons. Its translation and rotation components could be compared with the recoil and angular momentum of the remnant measured independently \cite{IozzoKheraEtAl:2021remnantComparison}.

Horizon shear is correlated with the gravitational-wave news, on the individual horizons during inspiral \cite{Prasad:2020news} and on the outer common horizon after it forms \cite{Prasad:2025shear}. In the ringdown regime, horizon tomography relates radiative fields at the horizon and at null infinity \cite{RibesMetidieriBongaKrishnan:2025tomography}. The common horizon forms inside the event horizon, so no signal from it can reach null infinity. A correlation is still possible, because the horizon and the outgoing radiation have a common source in the strong-field region. The formation law gives the geometry of the common horizon, and with it the shear, just after it forms. Comparing that shear with the news would test whether common horizon formation has a counterpart in the gravitational waveform. The formation law also bears on the onset of the ringdown. It is purely geometric, since it follows from how the time slices intersect a smooth tube, and it is unlikely to contain the ringdown of the remnant. The time at which the horizon departs from the formation law and begins to ring down \cite{Prasad:2025shear} would then show when the remnant can be described by black hole perturbation theory.

\begin{acknowledgments}
YCX thanks Abhay Ashtekar, Ivan Booth, Graham Cox, Peter Hintz, and B.~S.~Sathyaprakash for discussions. The National Science Foundation (NSF) supports YCX under Grant No.~\mbox{PHY-2608011} and VP under Grant Nos.~\mbox{PHY-2608388}, \mbox{AST-2307147}, \mbox{AST-2606673}, \mbox{PHY-2608011}, \mbox{PHY-2308886}, and \mbox{PHY-2309064}. YCX used Anvil at Purdue University \cite{anvil} and Bridges-2 at the Pittsburgh Supercomputing Center \cite{bridges2} through ACCESS allocations PHY260122, PHY260295, and PHY250294 \cite{access}, supported by NSF grants \#2138259, \#2138286, \#2138307, \#2137603, and \#2138296. OpenAI Codex and Claude Code assisted with the plotting code, and Claude Code assisted with language suggestions and polishing of this Letter.
\end{acknowledgments}

\noindent\textit{Data availability.} Processed data and analysis code are available on request; raw simulation outputs are unavailable because of their size.

\bibliography{refs}
\end{document}